\documentclass[10pt,conference]{IEEEtran}

\usepackage{cite}
\usepackage{amsmath,amssymb,amsfonts}
\usepackage{algorithmic}
\usepackage{graphicx}
\usepackage{textcomp}
\usepackage{xcolor}
\usepackage{booktabs} 
\usepackage{multirow}
\usepackage{xspace}
\usepackage{balance}
\usepackage{url}
\usepackage{caption}
\usepackage[para]{footmisc}
\usepackage{subcaption}
\usepackage[ruled,linesnumbered]{algorithm2e}
\usepackage[capitalize,noabbrev]{cleveref}
\usepackage[normalem]{ulem}
\usepackage{listings}
\usepackage{tabularx}
\usepackage{threeparttable}
\usepackage{tablefootnote}
\usepackage{colortbl}
\usepackage{tikz}
\usepackage{pgfplots}
\usepackage[many]{tcolorbox}
\pgfplotsset{width=10cm,compat=1.9}

\usepackage{amsthm}

\usepackage{booktabs}   
\usepackage{graphicx}   
\usepackage{enumitem}   

\newtcolorbox{boxH}{
    boxrule = 0pt, 
}

\definecolor{lightgray}{gray}{0.95}

\definecolor{highlight}{rgb}{0.67, 0.9, 0.7} 
\definecolor{codegreen}{rgb}{0,0.6,0}
\definecolor{codegray}{rgb}{0.5,0.5,0.5}
\definecolor{codepurple}{rgb}{0.58,0,0.82}
\definecolor{backcolour}{rgb}{0.95,0.95,0.92}

\lstdefinestyle{mystyle}{
    backgroundcolor=\color{backcolour},   
    commentstyle=\color{codegreen},
    keywordstyle=\color{magenta},
    numberstyle=\tiny\color{codegray},
    stringstyle=\color{codepurple},
    basicstyle=\ttfamily\footnotesize,
    breakatwhitespace=false,         
    breaklines=true,                 
    captionpos=b,                    
    keepspaces=true,                 
    numbers=left,                    
    numbersep=5pt,                  
    showspaces=false,                
    showstringspaces=false,
    showtabs=false,                  
    tabsize=2,
    escapeinside=||
}

\newcommand{\tch}{\operatorname{touch}}
\newcommand{\cov}{\operatorname{cover}}

\begin{document}

\title{

Assessing Behavioral Validation in UI Component Test Suites Using Inferred Metamorphic Relations
}
\author{\IEEEauthorblockN{Anonymous Author(s)}
}

\author{
\IEEEauthorblockN{
Yu Pei\IEEEauthorrefmark{1},
Cunming Zhang\IEEEauthorrefmark{1},
Jeongju Sohn\IEEEauthorrefmark{2},
and Mike Papadakis\IEEEauthorrefmark{1}
}

\IEEEauthorblockA{
\IEEEauthorrefmark{1}
University of Luxembourg\\
\{yu.pei, cunming.zhang, michail.papadakis\}@uni.lu
}

\IEEEauthorblockA{
\IEEEauthorrefmark{2}
Kyungpook National University\\
jeongju.sohn@knu.ac.kr
}
}

\maketitle
\IEEEpeerreviewmaketitle
\begin{abstract}

UI component libraries are commonly assessed using execution-based metrics such as statement and branch coverage, yet these metrics provide limited insight into whether tests verify the behavioral relations implied by component APIs and documentation. 
This paper presents an MR-based framework that uses inferred metamorphic relations (MRs) as an empirical behavioral reference, rather than a complete specification, for assessing UI component test suites. Given a component's source, documentation, and tests, the framework infers component-specific MRs using a UI-specific taxonomy, aligns tests with the inferred relations through hybrid deterministic and semantic analysis, and computes relation-level MR coverage metrics. We manually validate both the inferred MR space and the test--MR alignment.
Our evaluation shows that existing test suites exercise substantially more behavioral relations than they explicitly validate: MR \textit{Cover} remains between 42.5\% and 47.6\% across three LLM configurations and consistently below MR \textit{Touch}. Most uncovered relations are weak-oracle cases, where behaviors are exercised but lack explicit behavioral validation. MR coverage also complements execution-based coverage by revealing behavioral gaps not reflected by statement or branch coverage alone.
We further assess practical relevance through issue-description mapping, oracle strengthening, and MR-relevant injected faults. Most reported issue descriptions can be mapped to inferred MR relation types; weak-oracle relations often expose missing validation evidence; and MR labels show a trend in MR-relevant fault detection. Overall, MR coverage provides a complementary relation-level perspective for assessing behavioral validation in modern UI component testing.

\end{abstract}



\section{Introduction}\label{sec:introduciton}

UI component libraries such as material-ui, ant-design, and element-plus have become foundational infrastructure for modern front-end development~\cite{delgado2016reusing, abdalkareem2017developers, lazuardy2022modern}. They provide reusable interface foundations and are extensively employed in industrial and open-source ecosystems~\cite{decan2019empirical, naik2023awesome, weeraddana2024dependency}. Large open-source communities maintain popular libraries such as material-ui\footnote{https://mui.com/} and ant-design\footnote{https://ant.design/}, which are widely adopted across both industrial and open-source software ecosystems. Consequently, defects in a single component can affect a large number of downstream applications~\cite{venturini2023depended}. Unlike traditional utility libraries, however, UI components are defined not only by functional correctness but also by rich behavioral contracts governing interactions among component properties, user events, accessibility semantics, rendering behavior, and composition contexts~\cite{barr2014oracle,alshayban2020accessibility, zhou2025declarui, bajammal2021semantic}.

These behavioral contracts are only partially reflected by conventional execution-based coverage metrics. Statement and branch coverage indicate whether component code has been executed, but they provide limited insight into whether tests actually validate the behavioral relations that define component correctness~\cite{inozemtseva2014coverage,gopinath2014code,kochhar2017code,andrews2005mutation, schuler2011assessing}. A test may render a component, trigger user interactions, or execute the same control-flow path while failing to verify that disabled components suppress interaction, keyboard navigation preserves accessibility semantics, or responsive layouts maintain expected visual behavior. Consequently, high execution coverage does not necessarily imply high behavioral validation.

Metamorphic relations (MRs) provide a natural representation of such behavioral expectations~\cite{zhou2018metamorphic, zhang2023automated, li2025metamorphic}. Rather than specifying expected outputs for individual executions, an MR describes how observable behavior should change or remain invariant—under systematic transformations of component properties, interactions, state, or execution context~\cite{chen2018metamorphic,cho2025metamorphic,tsigkanos2023large}. MRs have been widely used to alleviate the oracle problem by generating follow-up tests or constructing relational test oracles in domains such as machine learning, scientific computing, and compiler testing~\cite{tian2018deeptest,zhang2018deeproad,chen2016empirical, NolascoMDGGPUAF24}. UI component APIs similarly imply behavioral relations across property changes, accessibility states, rendering contexts, and user interactions, suggesting that inferred MRs can serve not only as guidance for testing but also as an empirical behavioral reference for assessing existing test suites.

Despite these advances, existing work primarily uses MRs to generate tests or construct test oracles. Comparatively little attention has been paid to using inferred MRs as a behavioral specification for assessing the adequacy of existing UI component test suites. Existing execution-based metrics indicate whether component code is exercised, but provide limited evidence of whether tests explicitly validate the behavioral relations implied by component APIs and documentation~\cite{huo2014improving, goldstein2024property}. Consequently, developers still lack a practical way to measure the behavioral validation of UI component test suites beyond execution coverage.

This paper addresses this gap by presenting an MR-based framework for assessing the behavioral validation of UI component test suites. Rather than generating new tests, our framework uses inferred MRs as a behavioral reference space for evaluating existing tests. Given a component's implementation, public API specification, and test suite, the framework (1) infers a component-level MR space using a UI-specific taxonomy, (2) aligns existing tests with the inferred behavioral relations through hybrid deterministic and semantic analysis, and (3) distinguishes between \textit{behavioral reach} (\textit{Touch}) and \textit{behavioral validation} (\textit{Cover}). This relation-level perspective identifies whether behavioral gaps arise from missing test scenarios or from insufficient behavioral validation in tests that already exercise the relevant behavior.

While prior metamorphic testing work often uses MRs for follow-up test generation or oracle construction, we use inferred MRs as a behavioral reference for assessing the adequacy of existing UI component test suites.
We evaluate the framework on 214 components from four widely used UI component libraries. To assess the reliability of the inferred behavioral reference, we manually evaluate both MR inference and test--MR alignment: the inferred MRs achieve 88.6\% usability on a stratified sample, and hybrid alignment reaches an F1 of 96.6\% for \textit{Touch} and 89.7\% for \textit{Cover}. The results show that existing test suites exercise substantially more inferred behavioral relations than they explicitly validate, indicating that many behavioral relations are exercised but lack explicit behavioral validation. Furthermore, MR coverage provides a complementary perspective beyond statement and branch coverage by capturing behavioral validation not reflected by execution-based metrics. Finally, analyses of reported issue descriptions, an oracle-strengthening study, and an MR-relevant injected fault study support the practical relevance of MR coverage for assessing behavioral validation in existing UI component test suites.

In summary, the main contributions of this paper are:
\begin{itemize}
    \item \textbf{A UI-specific MR taxonomy.}
    We define a six-category taxonomy of component-level metamorphic relations covering input/prop, state/event semantics, interaction/accessibility, visual/layout, composition/context, and data flow.
    \item \textbf{A hybrid LLM-assisted framework for MR-based UI behavioral validation.}
    The framework combines taxonomy-constrained LLM inference with deterministic local-evidence alignment to determine which inferred behavioral relations are exercised (\textit{Touch}) and which are explicitly validated (\textit{Cover}) by existing tests.
    \item \textbf{An empirical study on 214 UI components.} 
    We evaluate the framework on 214 UI components and show that MR coverage complements execution-based coverage by distinguishing behavioral reach from behavioral validation.
    \item \textbf{Evidence of practical relevance.} Through analyses of reported issue descriptions, an oracle-strengthening study, and an MR-relevant injected fault study, we provide evidence that the inferred MR space captures behaviors reflected in real-world issues and that MR coverage offers a practically meaningful perspective for assessing behavioral validation.
\end{itemize}

\section{Background and Problem Statement}\label{sec:background}

\subsection{UI Component Behavior as Relational Contracts}\label{sec:background:components}
UI component libraries expose compact declarative APIs, but the behavior behind those APIs is often rich and stateful~\cite{jartarghar2022react, lu2025misty, liu2024enhancing}. A component configured through a small set of props, events, or slots may involve rendering logic, state-dependent branches, accessibility attributes, layout calculations, and callback dispatch~\cite{zhang2024llamatouch}. Consequently, correctness often depends on interactions among component properties, states, and execution contexts rather than on individual executions alone.
Many UI component behaviors are therefore inherently relational. Correctness is determined not only by the outcome of a single execution, but also by how observable outputs change under variations in inputs, state, interaction, or context. For example, changing a placement option should consistently update the rendered position, while keyboard navigation should preserve focus and accessibility semantics. Assessing UI component correctness therefore requires reasoning about behavioral relations across executions rather than validating each execution in isolation.

\subsection{Execution Coverage and the Oracle Gap}\label{sec:background:coverage}
UI component libraries are commonly tested using rendering frameworks, interaction utilities, and coverage instrumentation tools. 
Existing UI component test suites commonly exercise behaviors through rendering, state updates, and simulated user interactions. However, exercising a behavior does not necessarily imply that the expected behavioral relation is explicitly validated.

Execution-based metrics, such as statement and branch coverage, indicate whether implementation code is exercised during testing, but provide limited evidence about whether the exercised behavior is behaviorally validated. A test that merely executes a branch and a test that explicitly verifies the behavioral relation implemented by that branch contribute equally to structural coverage. This limitation gives rise to an oracle gap: behavioral relations may be exercised without being explicitly checked. Consequently, high execution coverage alone does not necessarily indicate strong behavioral validation.


\subsection{Metamorphic Relations as an Adequacy Basis}\label{sec:background:mt}

The oracle gap arises because UI component correctness is often defined by relationships among executions rather than by the outcome of a single execution. Metamorphic relations (MRs) naturally represent such expectations by specifying how observable outputs should relate across executions whose inputs or execution contexts are systematically transformed.
For UI components, MRs capture behavioral expectations such as consistency, monotonicity, invariance, and accessibility-preserving transformations. Although these relations are often implicit in component specifications and developer expectations, they provide a useful behavioral reference for assessing existing test suites.

Rather than asking only which implementation code is executed, we can ask which inferred behavioral relations are exercised and explicitly validated. This motivates the central question of this work: can inferred component-level MRs serve as an empirical behavioral reference for assessing the behavioral adequacy of existing UI component test suites? 
\section{Approach}\label{sec:approach}

\begin{figure*}[h]
  \centering
  \includegraphics[width=0.96\linewidth]{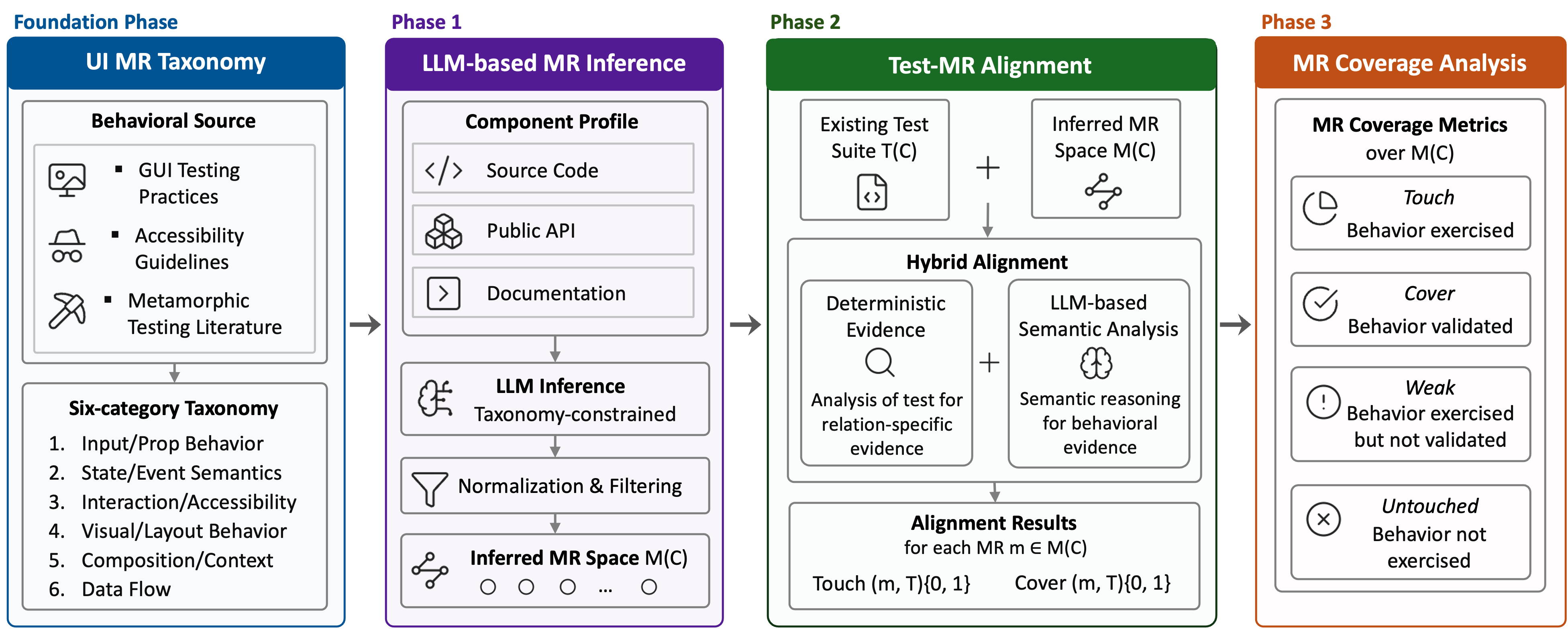}
  \caption{Overview of the MR-based behavioral validation framework.}
  \label{fig:workflow}
\end{figure*}

\subsection{Overview}\label{sec:approach:overview}
~\cref{fig:workflow} presents the overall framework. Given a UI component $C$ and its existing test suite $T(C)$, the framework first uses a UI MR taxonomy to construct a component-specific MR space $M(C)$ from the component's public API, documentation, and implementation. It then aligns the inferred MRs with existing tests to determine which behavioral relations are exercised and which are explicitly validated. Finally, it computes MR-based completeness metrics that characterize behavioral reach and validation in the test suite. These metrics provide the basis for the empirical analyses in Section~\ref{sec:result_analysis}.

\subsection{UI MR Taxonomy}\label{sec:approach:taxonomy}

To provide a structured representation of UI component behavior, we derive a UI-specific MR taxonomy by synthesizing behavioral dimensions commonly reflected in component APIs, official documentation, GUI testing, accessibility guidelines, and metamorphic testing. We consolidate related behavioral concepts into six top-level categories: input/prop behavior, state/event semantics, interaction/accessibility, visual/layout behavior, composition/context behavior, and data flow. The \textit{Foundation} phase in \cref{fig:workflow} summarizes the taxonomy.
The taxonomy constrains MR inference to predefined behavioral relation types while allowing component-specific instantiations. This design reduces arbitrary relation generation and promotes more consistent MR inference across component libraries.

\begin{figure}[h]
  \centering
  \scalebox{1.03}{\includegraphics[width=\linewidth]{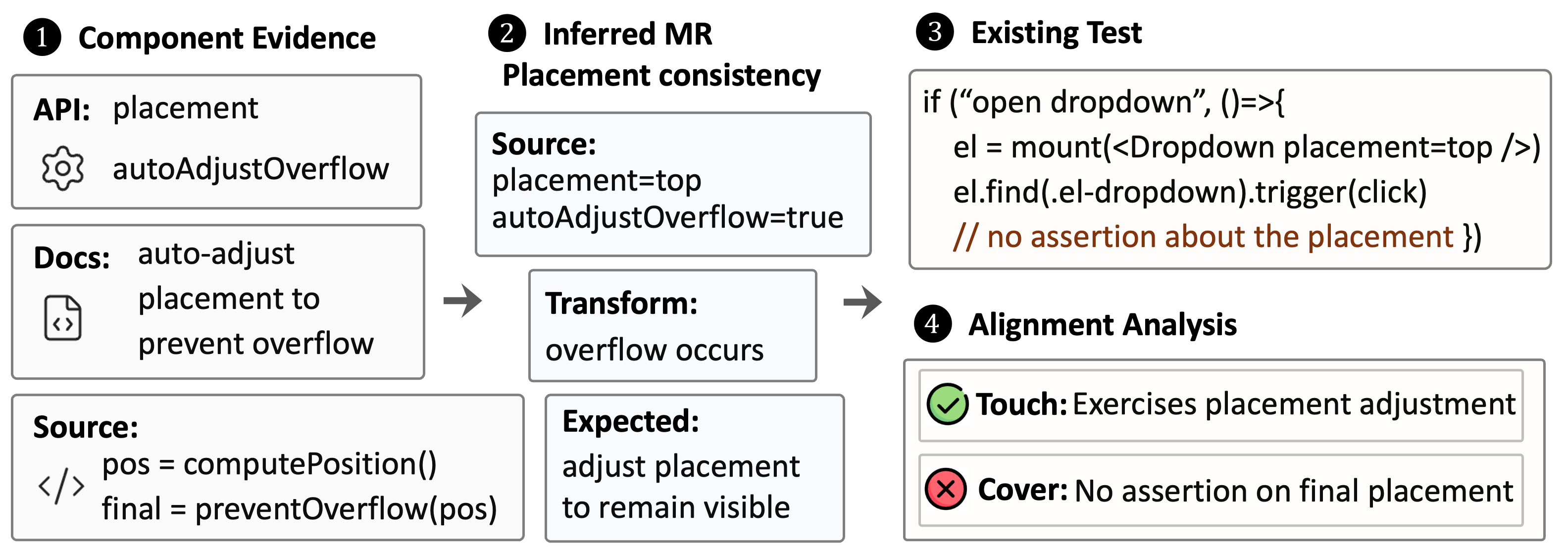}}
  \caption{Example: placement consistency MR is inferred from the dropdown component. The existing test exercises the auto-adjust behavior (\textit{Touch}) but does not assert the rendered placement, so the relation is not \textit{Cover}.}
  \label{fig:inferexample}
\end{figure}

To make the subsequent inference and alignment steps concrete, \cref{fig:inferexample} presents a dropdown example. It illustrates how component evidence is used to infer a placement-consistency MR and how an existing test is later aligned with that relation. In this example, the test exercises auto-adjust placement behavior but does not assert the rendered placement outcome, forming a weak-oracle case that is revisited in Phase~1 and Phase~2.

\subsection{Phase 1: LLM-based MR Inference}\label{sec:approach:mr-construction}

For each component, we first construct a component profile by extracting information from its public API, documentation, and implementation. The profile summarizes observable behavioral information, including configurable properties, documented behavioral expectations, state transitions, event handling, rendering logic, and accessibility semantics. Conditioned on the component profile and the predefined UI MR taxonomy, an LLM instantiates component-specific metamorphic relations. The LLM is constrained by the predefined taxonomy and required to generate structured MR descriptions consisting of a relation type, expected behavioral relation, and supporting evidence.

Each inferred MR consists of (1) a source condition, (2) an input, interaction, or state transformation, (3) an expected behavioral relation, and (4) supporting evidence extracted from the component artifacts.
As shown in \cref{fig:inferexample}, the placement consistency MR is inferred from evidence in the component profile, including placement-related API options, documentation, and implementation logic. The resulting relation specifies the source condition, transformation, expected behavioral relation, and supporting evidence used in later alignment.

The inferred relations are subsequently normalized and filtered. Relations relying on undocumented behavior, unobservable outputs, or inconsistent taxonomy labels are removed, and semantically equivalent relations are merged. The resulting MR set, denoted by $M(C)$, serves as the behavioral reference space for the subsequent completeness analysis. Its validity and taxonomy consistency are evaluated in Section~\ref{sec:validation}.

\subsection{Phase 2: Test--MR Coverage Alignment}\label{sec:approach:alignment}
Given the inferred MR space $M(C)$ and the existing test suite $T(C)$, we determine how each test relates to each inferred MR. We distinguish between two levels of behavioral validation. A test \textit{touches} an MR if it exercises the behavioral condition described by that relation. A test \textit{covers} an MR only if it additionally contains an oracle that explicitly validates the expected behavioral relation. Accordingly, we define two metrics:
\[
touch(m,T) \in \{0,1\}, \qquad cover(m,T) \in \{0,1\}.
\]

By definition, every covered relation is also touched; that is,
\[
cover(m,T)=1 \Rightarrow touch(m,T)=1.
\]

We compute these metrics using a hybrid alignment procedure that combines deterministic evidence extracted from test code with LLM-based semantic analysis. Following a disjunctive fusion strategy, an MR is labeled as \textit{Touch} or \textit{Cover} whenever either source provides supporting evidence. For \textit{Cover}, we require evidence that the test explicitly verifies the expected behavioral relation. An MR is therefore not labeled as \textit{Cover} merely because the test contains an assertion; the assertion must be relation-relevant, i.e., it must check an observable outcome implied by the inferred MR rather than simply mention or execute the corresponding behavior. Because disjunctive fusion may over-approximate alignment labels, we evaluate its precision--recall trade-off in Section~\ref{sec:validation}.

The deterministic analysis parses the test suite into individual \textit{it()}/\textit{test()} blocks, where each block defines one test case in JavaScript/TypeScript testing frameworks and searches for predefined evidence patterns relevant to the inferred relation. Such evidence is derived from the MR description and relation type and includes component configurations, event triggers, user interactions, DOM queries, callback invocations, and assertion statements.
A \textit{Touch} decision requires evidence that the behavioral condition described by the MR is exercised. A \textit{Cover} decision is more restrictive and requires evidence of (1) a relation-relevant interaction or transformation and (2) an explicit assertion validating the expected behavioral relation within the same test block. Typical oracle patterns include assertions expressed through \textit{expect()}, \textit{assert()}, or equivalent matcher APIs. The LLM-based semantic analysis applies the same behavioral criteria but reasons over the test at a semantic level, allowing it to recognize indirect oracle patterns that cannot be reliably identified through deterministic matching alone. 

For reporting and validation, alignment decisions are attributed to one of two categories. A decision is considered \textit{deterministic-supported} whenever deterministic analysis identifies supporting evidence, regardless of whether the LLM reaches the same conclusion. Otherwise, the decision is categorized as \textit{semantic-only}. 
We evaluate the reliability of the resulting alignment decisions in Section~\ref{sec:validation}. 
The output of this phase is an MR--test alignment that determines which inferred behavioral relations are exercised (\textit{Touch}) and which are explicitly validated (\textit{Cover}), providing the basis for the completeness metrics introduced in the next phase.

\subsection{Phase 3: MR Coverage Analysis}\label{sec:approach:completeness}
Using the alignment results, we compute component-level MR coverage metrics over the MR space $M(C)$. \textit{Touch} measures the proportion of inferred behavioral relations exercised by the test suite:
{\small
\[
\mathit{Touch} = \frac{|\{ m \in M(C) \mid \tch(m,T) \}|}{|M(C)|}.
\] 
}

\textit{Cover} measures the proportion of relations that are both exercised and explicitly validated:
{\small
\[
\mathit{Cover} = \frac{|\{ m \in M(C) \mid \cov(m,T) \}|}{|M(C)|}
\] 
}
To characterize behavioral gaps, we further distinguish two categories. An \emph{Untouched} relation is not exercised by any test:
{\small
\[
\mathit{Untouched} = \frac{|\{ m \in M(C) \mid \neg \tch(m,T) \}|}{|M(C)|}
\]}

A \emph{Weak} relation is exercised but lacks validation of the expected behavioral relation:
{\small
\[
\mathit{Weak}=\frac{
  \bigl| \{ m \in M(C) \mid \tch(m,T) \land \neg \cov(m,T) \} \bigr|
}{|M(C)|}
\]
}

These metrics distinguish missing behavioral scenarios from insufficient behavioral validation and support the analyses reported in Section~\ref{sec:result_analysis}.

\section{Evaluation Design}\label{sec:evaluation}
Since inferred MRs are not ground-truth specifications, we evaluate whether they can serve as a reliable behavioral reference for aggregate adequacy analysis. We assess relation-level coverage, gap characteristics, complementarity with execution-based coverage, and practical relevance.

\subsection{Research Questions}\label{sec:eva:rqs}
Our evaluation is organized around four research questions.

\textbf{RQ1 (MR-based Behavioral Coverage):} To what extent do existing UI component test suites exercise (\textit{Touch}) and explicitly validate (\textit{Cover}) the inferred behavioral relations?

\textbf{RQ2 (Behavioral Gaps):} Which behavioral relation types are most often left \textit{Untouched} or \textit{Weak}, and what do these gaps suggest about existing UI component tests?

\textbf{RQ3 (Complementary Adequacy Signals):} Does MR coverage provide adequacy information beyond statement and branch coverage?

\textbf{RQ4 (Practical Relevance):} Are reported issue descriptions represented in the inferred MR space, and do MR-based labels provide supporting evidence for practical behavioral validation?

\subsection{Dataset}\label{sec:dataset}
We evaluate the framework on four widely used open-source UI component libraries: \textit{ant-design}, \textit{element-plus}, \textit{material-ui}, and \textit{base-ui}. These libraries represent mature, production-grade UI ecosystems spanning both React and Vue, with diverse implementation strategies, documentation styles, and testing practices. Together, they provide a diverse benchmark for evaluating behavioral validation across modern component libraries. 
For each component, we collect its implementation, public documentation/API specification, and test suite. These artifacts provide the inputs required to construct component profiles, infer behavioral MRs, and align inferred relations with existing tests.

Table~\ref{tab:dataset} summarizes the dataset. In total, it contains 214 UI components distributed across five common categories: Inputs, Data Display, Navigation, Feedback, and Layout. The category labels follow the official classifications used by the libraries studied and are used only for descriptive analyses. All MR inference, alignment, and MR coverage measurements are performed at the individual component level.

\begin{table}[h]
\centering
\caption{Studied libraries, components, and their classification. \#Comp. denotes the number of components. $LOC$ denotes average lines of code per component.}
\label{tab:dataset}
\small
\setlength{\tabcolsep}{2pt}
\begin{tabular}{lrrrrrr}
\toprule
&  & & \textbf{ } & & \multicolumn{2}{c}{\textbf{Comp.} ($LOC$)} \\
\textbf{Library} & \textbf{Stars} & \textbf{Commits} &  \textbf{Issues$^\dagger$} & \textbf{\#Comp.} &  \textbf{Src} & \textbf{Test}  \\
\midrule
ant-design\footnotemark[1]   & 98.2k & 32{,}243 & 1{,}224 & 63 & 653   & 393  \\
element-plus\footnotemark[2] & 27.5k &  7{,}636 &   891   & 71 & 736   & 523  \\
material-ui\footnotemark[3] & 98.4k & 28{,}304 & 1{,}399 & 47 & 488   & 452  \\
base-ui\footnotemark[4]  &  9.8k &  4{,}362 &   316   & 33 & 1{,}213 & 1{,}643 \\
\midrule
\textbf{Total} & & & & \textbf{214}  & & \\
\bottomrule
\multicolumn{7}{l}{\footnotesize $^\dagger$Repository and open issues were collected as of 2026-05-10.}
\end{tabular}
\vspace{6pt}

\begin{tabular}{@{}lrrrrr@{}}
\toprule
\textbf{Category} & \textbf{Inputs} & \textbf{Data display} & \textbf{Navigation} & \textbf{Feedback} & \textbf{Layout} \\
\midrule
\#Comp.   & 69 & 56 & 36 & 34 & 19 \\
\bottomrule
\end{tabular}
\end{table}
\footnotetext[1]{\url{https://github.com/ant-design/ant-design}}
\footnotetext[2]{\url{https://github.com/element-plus/element-plus}}
\footnotetext[3]{\url{https://github.com/mui/material-ui}}
\footnotetext[4]{\url{https://github.com/mui/base-ui}}

\subsection{Measurement Validation}\label{sec:validation}
All subsequent analyses rely on the inferred MR space and the resulting MR coverage measurements. We therefore validate both MR inference and test--MR alignment, since errors in either step can affect \textit{Touch} and \textit{Cover}. All LLM-assisted inference and semantic alignment use fixed prompts, identical procedures, and a temperature of 0.2 across components.

\paragraph{MR validity and taxonomy consistency.}
We randomly sample 420 inferred MRs (approximately 10\% of the total) using stratified sampling across taxonomy categories, libraries, and component types. Two authors independently assess whether each MR is supported by the component specification, observable through component behavior, testable in principle, non-duplicative, and correctly classified within the proposed taxonomy. Disagreements are resolved through discussion and subsequently reviewed by an experienced front-end developer. 
Manual validation showed that among the sampled MRs, 88.6\% are judged usable after manual inspection, while 6.4\% are considered invalid or unsupported, and 5.0\% are identified as duplicates. The taxonomy achieves agreement of 94.8\% at the top level and 89.5\% at the fine-grained relation level. These results provide supporting evidence that the inferred MR space and taxonomy are sufficiently reliable for the subsequent MR coverage analyses.



\paragraph{Alignment Validation.}
Since MR coverage depends on test--MR alignment, we evaluate the reliability of the alignment procedure and the contribution of its two evidence channels. For each LLM configuration, we sampled 100 MR--test instances from each of four groups: \textit{Touch}-positive, \textit{Touch}-negative, \textit{Cover}-positive, and \textit{Cover}-negative. After removing duplicates, the validation sets contained 386, 385, and 387 instances for DeepSeek, Gemini, and GPT, respectively. Each sampled instance was reviewed by one author. Cases with unclear evidence were independently reviewed by a second author until consensus was reached. We did not compute an inter-rater agreement statistic because the second review was performed as an adjudication step rather than an independent annotation process. We compare deterministic-only, semantic-only, and hybrid alignment, and report precision (P), recall (R), and F1 averaged over the three LLM configurations.

Table~\ref{tab:alignment-ablation} summarizes the results. Deterministic-only alignment performs well for \textit{Touch}, but has substantially lower \textit{Cover} recall because explicit behavioral validation is harder to identify than local test actions. Semantic-only alignment achieves the highest precision but lower recall, indicating that LLM-only judgments are more conservative on this sample. The hybrid variant achieves the highest F1 for both \textit{Touch} and \textit{Cover}, with no observed false negatives in the sampled validation set. Because the validation set is stratified over predicted labels, these results should be interpreted as sample-level reliability evidence rather than population-level estimates. We therefore use hybrid alignment in the main evaluation, while treating the resulting labels as aggregate analysis signals rather than ground truth.

\begin{table}[h]
\centering
\caption{Ablation of alignment evidence across LLM configurations. Results report average precision, recall, and F1 over three LLMs on manually validated MR--test instances.}
\label{tab:alignment-ablation}
\small
\setlength{\tabcolsep}{2pt}
\begin{tabular}{lcccccc}
\toprule
& \multicolumn{3}{c}{\textbf{Touch} (\%)}& \multicolumn{3}{c}{\textbf{Cover} (\%)} \\
 \cmidrule(lr){2-4} \cmidrule(lr){5-7}
\textbf{Variant} & Precision & Recall & F1 & Precision & Recall & F1 \\
\midrule
Deterministic-only
& 92.9 & 90.4 & 91.5
& 72.6 & 50.7 & 58.9 \\
Semantic-only
& 100.0 & 73.1 & 84.2
& 94.1 & 71.8 & 81.1 \\
Hybrid
& 93.4 & 100.0 & 96.6
& 81.6 & 100.0 & 89.7 \\
\bottomrule
\end{tabular}
\end{table}



\subsection{Execution-based Coverage Baselines}\label{sec:baselines}
We compare MR coverage with statement and branch coverage as conventional execution-based baselines. For each library, we run the original test suite using its native infrastructure and collect coverage over the same component scope used for MR analysis. At the component level, \textit{Stmt} and \textit{Branch} measure code execution, while \textit{Touch} and \textit{Cover} measure behavioral reach and behavioral validation.

\subsection{Practical Relevance Assessment}\label{sec:external-validation}
We assess external relevance from three complementary perspectives: issue-related alignment, oracle augmentation for weak relations, and detection of MR-relevant injected faults.

\paragraph{Issue-based validation.} To assess the external relevance of the inferred MR space, we examine whether behaviors described in real-world issue reports can be represented by inferred MR relation types. We collected component-related GitHub issues and pull requests from the four studied libraries. Reports unrelated to component behavior (e.g., build or infrastructure issues) were excluded through manual inspection before annotation. We then report the mapping rate and analyze how mapped issue descriptions are distributed across relation types with different average MR \textit{Cover} levels. The average MR \textit{Cover} for each relation type is computed from the DeepSeek-derived MR space and corresponding hybrid-alignment results. This analysis is intended as an external relevance assessment rather than causal evidence: issue reports are influenced by project popularity, reporting practices, and maintenance processes. We therefore do not interpret issue frequency as defect proneness, but use it to assess whether reported behaviors are represented in the inferred MR space and whether MR \textit{Cover} provides additional context for understanding their behavioral validation.

\paragraph{Weak-oracle assertion study.} To assess the practical relevance of identified weak-oracle gaps, we conduct a targeted oracle-strengthening study. We sample 200 weak-oracle relations across MR types and two libraries (\textit{ant-design} and \textit{element-plus}) with reproducible test environments. For each relation, we augment the corresponding test block with a single relation-specific assertion derived from the inferred MR. Assertion templates are instantiated using behavioral cues from the MR (e.g., placement or callback identifiers), or a conservative default assertion when no concrete cue is available. Two authors inspect failing executions and separate behavioral inconsistencies from snapshot, infrastructure, or environmental artifacts.

\paragraph{MR-relevant injected fault study.}
To assess whether MR coverage labels are associated with detecting MR-relevant behavioral faults, we conduct an MR-relevant injected fault study on 200 sampled MRs: 80 \textit{Covered}, 80 \textit{Weak}, and 40 \textit{Untouched}, stratified by MR type within each status.
For each MR, we inject one relation-specific behavioral fault into the component implementation, targeting behaviors such as disabled-state handling, ARIA mappings, state synchronization, placement behavior, or event-order semantics. We then execute the original component test suite and compare detection rates across \textit{Covered}, \textit{Weak}, and \textit{Untouched} relations.
\section{Results}\label{sec:result_analysis}

\subsection{RQ1: MR-based Behavioral Coverage}\label{sec:results:rq1}
We first examine the gap between behavioral reach and behavioral validation in existing component test suites.
Table~\ref{tab:completeness} reports MR completeness across libraries and component categories using the \textit{Touch} and \textit{Cover} metrics.

\subsubsection{Existing tests achieve high behavioral reach but substantially lower behavioral validation.} 
Table~\ref{tab:completeness} shows that \textit{Touch} is consistently higher than \textit{Cover} across libraries, categories, and LLM configurations\footnote{DeepSeek (\textit{deepseek-chat}), Gemini (\textit{gemini-2.5-flash}), and GPT (\textit{gpt-5-mini}).}. At the model level, the overall MR \textit{Cover} rate ranges from 42.5\% to 47.6\% across the three LLM configurations, while MR \textit{Touch} remains substantially higher.
Using \textit{Gemini}, library-level \textit{Touch} ranges from 85.4\% to 89.4\%, whereas \textit{Cover} ranges from 41.9\% to 54.7\%. This gap suggests that existing tests often exercise the inputs, states, or interactions associated with inferred MRs, but do not always include assertions that validate the expected behavioral relation. Thus, within the inferred MR space, many uncovered relations arise not because the associated behavior is absent from tests, but because the tests do not provide explicit relation-level evidence for the expected behavior.

\begin{table}[h]
\centering
\caption{MR completeness by UI library and component category.}
\label{tab:completeness}
\setlength{\tabcolsep}{2pt}
\begin{tabular}{l l ccc ccc}
\toprule
 &  & \multicolumn{3}{c}{\textbf{Touch} (\%)} & \multicolumn{3}{c}{\textbf{Cover} (\%)} \\
\cmidrule(lr){3-5} \cmidrule(lr){6-8}
\textbf{Group} & \textbf{Item} & Gemini & DeepSeek & GPT & Gemini & DeepSeek & GPT \\
\midrule
\multirow{4}{*}{Library} 
 &  ant-design                           & 88.2 & 83.0 & 82.7 & 45.0 & 40.9 & 38.1 \\
 & element-plus                        & 85.4 & 78.3 & 81.2 & 54.7 & 50.3 & 50.5 \\
 &  base-ui                             & 89.4 & 84.4 & 86.8 & 45.7 & 43.7 & 40.8 \\
 & material-ui                         & 88.6 & 78.0 & 83.1 & 41.9 & 41.5 & 37.6 \\
\midrule
\multirow{5}{*}{Category}
 &  Data display                        & 84.8 & 74.5 & 79.4 & 40.1 & 36.5 & 36.1 \\
 & Feedback                            & 88.1 & 80.4 & 80.6 & 45.6 & 44.9 & 36.4 \\
 &  Inputs                               & 89.3 & 87.9 & 88.9 & 57.8 & 56.3 & 55.9 \\
 & Layout                              & 76.0 & 60.2 & 67.2 & 37.6 & 22.7 & 25.2 \\
 &  Navigation                           & 93.7 & 86.8 & 87.3 & 47.1 & 46.0 & 41.5 \\
\bottomrule
\end{tabular}
\end{table}

\subsubsection{Completeness varies across libraries and component categories.} At the library level, \textit{element-plus} shows the highest \textit{Cover}, exceeding 50\% across all three models, whereas \textit{material-ui} remains between 37.6\% and 41.9\%. These differences may reflect variation in library-specific testing practices, component designs, or assertion styles.
Variation is also visible across component categories. \textit{Input} components achieve the highest \textit{Cover} of 57.8\%, whereas \textit{Layout} components achieve the lowest (22.7--37.6\%). This pattern is consistent with the nature of the underlying behaviors: input-related relations often expose explicit state transitions, validation outcomes, and callback effects that are straightforward to assert, while layout relations more often involve visual, spatial, or responsive behavior that may be harder to validate through conventional component tests.

\begin{tcolorbox}[colback=gray!5, colframe=gray!30, title=\textbf{\small Answer to RQ1}, coltitle=black!90, left=2.5pt, top=3pt, bottom=3pt, right=2.5pt, boxsep=1.8pt]
Existing UI component test suites achieve high behavioral reach but lower behavioral validation. Across libraries and models, inferred behavioral relations are more often exercised than covered. Validation strength also varies by component category, with \textit{Input} components consistently better covered than \textit{Layout} components.
\end{tcolorbox}

\subsection{RQ2: Behavioral Gaps}\label{sec:rq2-results}

RQ2 examines where behavioral gaps arise among inferred relations that are not covered by existing tests. We first decompose uncovered relations into \textit{Weak} and \textit{Untouched} relations, and then identify the relation types that remain consistently under-covered across LLMs.

\subsubsection{Most uncovered relations are associated with weak validation rather than missing test scenarios.}

\begin{figure}[h]
  \centering
  \includegraphics[width=0.86\linewidth]{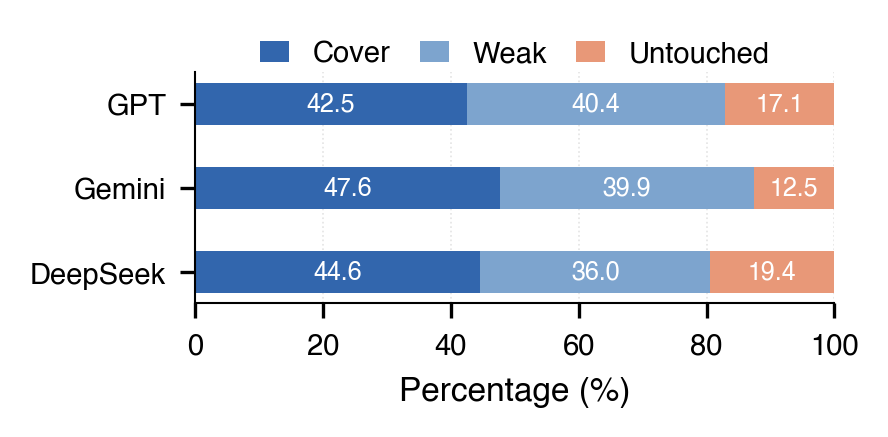}
  \caption{Distribution of MR coverage outcomes (Cover, Weak, Untouched) for three LLMs across 214 components.}
  \label{fig:rq2_gap_decomp}
\end{figure}

Figure~\ref{fig:rq2_gap_decomp} divides inferred relations into \textit{Cover}, \textit{Weak}, and \textit{Untouched}. Across the three LLMs, 52.4--57.5\% of inferred relations remain uncovered under the \textit{Cover} criterion. Weak-oracle cases form the majority of these uncovered relations, accounting for roughly two-thirds to three-quarters of them. In contrast, only 12.5--19.4\% of all inferred relations remain completely untouched. This indicates that most uncovered relations are exercised by existing tests but are not explicitly validated. Thus, the observed gaps are more often associated with limited behavioral validation than with entirely missing test scenarios.

\begin{figure}[h]
  \centering
  \includegraphics[width=0.96\linewidth]{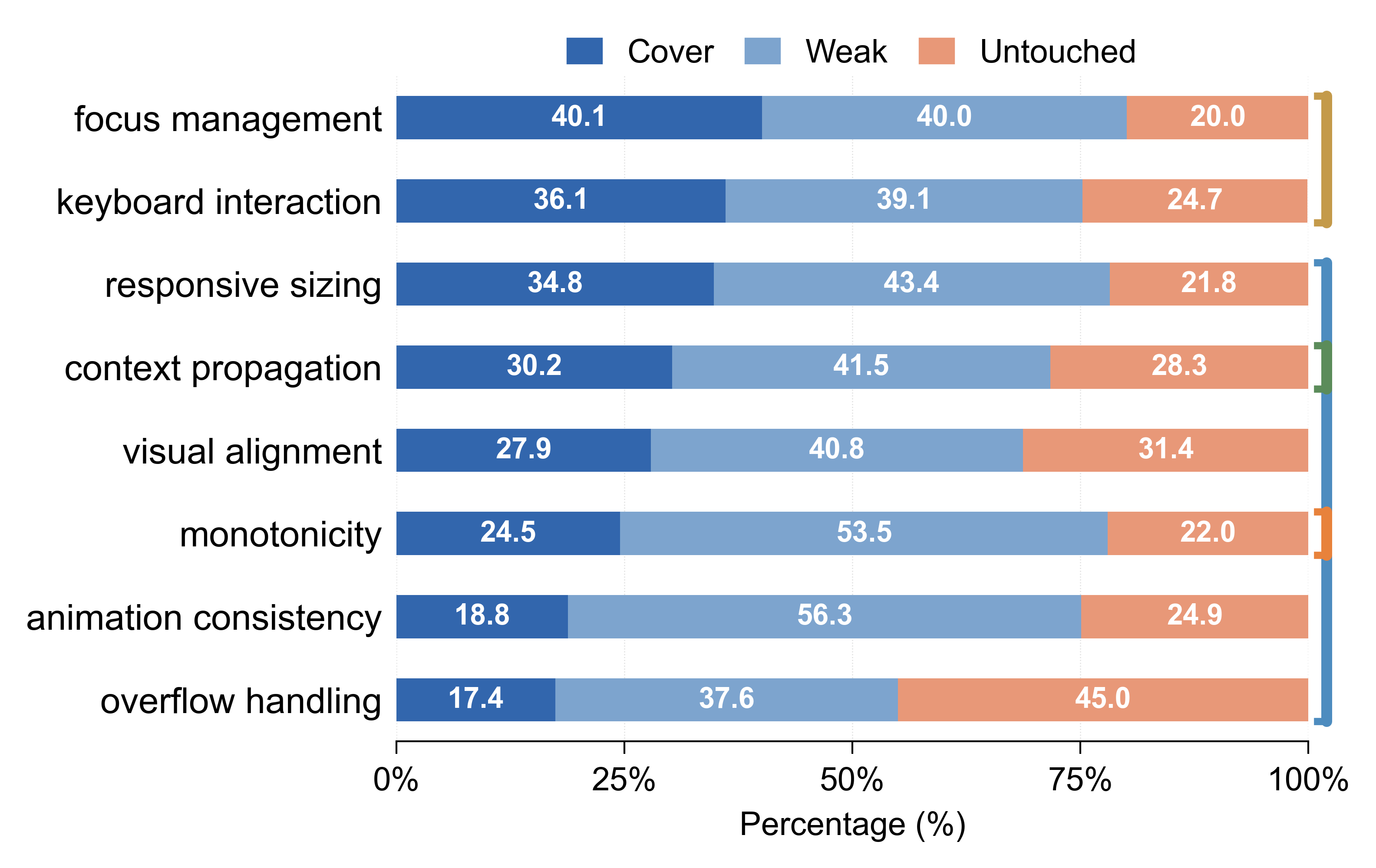}
  \caption{Relation types with the lowest \textit{Cover} rates, grouped by top-level categories. Colored brackets distinguish the groups (blue: Visual/layout, orange: Input/prop, green: Composition/\
  context, brown: Interaction/accessibility).}
  \label{fig:rq2_worst_types}
\end{figure}

\subsubsection{Visual/layout behavior is prominent among the lowest-covered relation types.} 

Figure~\ref{fig:rq2_worst_types} shows the eight relation types with the lowest \textit{Cover} rates. Visual/layout behavior is the most represented top-level category in this group, covering four of the eight relation types: \textit{overflow handling} (17.4\%), \textit{animation consistency} (18.8\%), \textit{visual alignment} (27.9\%), and \textit{responsive sizing} (34.8\%).
The breakdown further shows that low \textit{Cover} can arise from different sources. For \textit{overflow handling}, the gap is mainly due to missing behavioral reach: 45.0\% of inferred relations are \textit{Untouched}, suggesting that overflow, clipping, and boundary-layout behaviors are often absent from existing test scenarios. In contrast, \textit{animation consistency} and \textit{monotonicity} have high \textit{Weak} rates, at 56.3\% and 53.5\%, respectively. These relations are often exercised but not explicitly validated, indicating a weak-oracle gap rather than a complete absence of test scenarios.

Similar weak-oracle patterns also appear in \textit{context propagation}, \textit{keyboard interaction}, and \textit{focus management}. These behaviors often require assertions over layout changes, visual state, accessibility attributes, focus state, or cross-component interactions, which are less directly captured by conventional rendering or state-transition checks.

\begin{tcolorbox}[colback=gray!5, colframe=gray!30, title=\textbf{\small Answer to RQ2}, coltitle=black!90, left=2.5pt, top=3pt, bottom=3pt, right=2.5pt, boxsep=1.8pt]
MR gaps are dominated by weak validation rather than entirely missing behavioral scenarios. Across LLMs, weak-oracle relations form the majority of uncovered relations, suggesting that existing tests often exercise relevant behaviors without explicitly checking the expected relations. Low-coverage relation types are frequently observed in visual/layout, accessibility, and contextual behaviors.
\end{tcolorbox}

\subsection{RQ3: Complementary Adequacy Signals}\label{sec:rq3-results}
RQ3 investigates whether MR coverage provides a relation-level adequacy signal beyond statement and branch coverage. We analyze this relationship using the 201 components for which runtime coverage data are available.

\subsubsection{MR Coverage Changes with Sampled Test-suite Size.}
To examine how MR-based metrics change as test suites grow, we construct sampled sub-suites containing 25\%, 50\%, and 75\% of the original tests, repeat the sampling 10 times for each size, and compare them with the full test suites. Table~\ref{tab:rq3-merged} shows that all metrics increase with sampled test-suite size. Statement coverage rises from 54.3\% to 92.7\%, branch coverage from 61.0\% to 83.5\%, MR \textit{Touch} from 44.8\% to 82.2\%, and MR \textit{Cover} from 18.8\% to 45.2\%. Thus, MR coverage grows with test-suite size rather than being detached from test execution.

Statement coverage is moderately correlated with MR \textit{Touch} for smaller sampled suites (e.g., $\rho=0.41$ at 25\%), which is expected because \textit{Touch} only requires exercising the corresponding behavior. In contrast, correlations involving MR \textit{Cover} remain weak and approach zero for the full test suites. Thus, although both metric families increase with test-suite size, structural coverage provides only a partial view of behavioral adequacy: MR \textit{Cover} reflects whether exercised behaviors are explicitly validated, not merely executed.

\begin{table}[h]
\centering
\caption{Coverage growth and within-size Spearman correlations under sampled test-suite sizes ($n=201$ components; 10 random samples per size).}
\label{tab:rq3-merged}
\small
\setlength{\tabcolsep}{4pt}
\begin{tabular}{lrrrr}
\toprule
\textbf{Metric} & \textbf{25\%} & \textbf{50\%} & \textbf{75\%} & \textbf{100\%} \\
\midrule
\multicolumn{5}{l}{\textbf{Mean coverage metrics}} \\
Statement coverage (\%) & 54.3 & 68.9 & 78.7 & 92.7 \\
Branch coverage (\%)    & 61.0 & 69.7 & 75.8 & 83.5 \\
MR Touch (\%)           & 44.8 & 58.6 & 66.6 & 82.2 \\
MR Cover (\%)           & 18.8 & 28.1 & 34.5 & 45.2 \\
\midrule
\multicolumn{5}{l}{\textbf{Within-size Spearman correlations}} \\
$\rho$(Statement, MR Cover) & 0.24 & 0.19 & 0.12 & $-$0.16 \\
$\rho$(Branch, MR Cover)    & 0.00 & 0.03 & 0.01 & $-$0.07 \\
$\rho$(Statement, MR Touch) & 0.41 & 0.30 & 0.19 & $-$0.12 \\
$\rho$(Branch, MR Touch)    & 0.07 & 0.08 & 0.05 & 0.02 \\
\bottomrule
\end{tabular}
\end{table}

\subsubsection{Execution Coverage Provides Limited Evidence of Behavioral Validation.} 
We further examine whether the conditional-probability results are sensitive to the \textit{MR Cover} threshold by repeating the analysis with thresholds of 40\%, 50\%, and 60\%. As shown in Table~\ref{tab:rq3-conditional}, the same asymmetry is observed across thresholds. Among components with high statement coverage, the proportion that also reaches the \textit{MR Cover} threshold decreases from 58.2\% to 28.8\% as the threshold becomes stricter. A similar pattern appears for high branch coverage, where the proportion decreases from 61.0\% to 30.1\%. In contrast, components that reach the \textit{MR Cover} threshold more often also have high statement or branch coverage. These results suggest that high \textit{MR Cover} often co-occurs with high execution coverage, but high execution coverage alone does not necessarily indicate high behavioral validation.

\begin{table}[h]
\centering
\caption{Sensitivity of conditional probabilities under different \textit{MR Cover} thresholds. $\mathrm{MRC}_{\theta}$ denotes MR \textit{Cover} $\geq \theta$. Percentages are shown with counts in parentheses.}
\label{tab:rq3-conditional}
\setlength{\tabcolsep}{1.5pt}
\renewcommand{\arraystretch}{1.3}
\begin{tabular}{lrrr}
\toprule
\textbf{Condition} & \textbf{$\theta=40\%$} & \textbf{$\theta=50\%$} & \textbf{$\theta=60\%$} \\
\midrule
$P(\mathrm{MRC}_{\theta} \mid \mathrm{Stmt}\geq90\%)$
& 58.2 {\footnotesize (85/146)}
& 41.8 {\footnotesize (61/146)}
& 28.8 {\footnotesize (42/146)} \\

$P(\mathrm{Stmt}\geq90\% \mid \mathrm{MRC}_{\theta})$
& 73.3 {\footnotesize (85/116)}
& 70.1 {\footnotesize (61/87)}
& 65.6 {\footnotesize (42/64)} \\

$P(\mathrm{MRC}_{\theta} \mid \mathrm{Branch}\geq80\%)$
& 61.0 {\footnotesize (83/136)}
& 43.4 {\footnotesize (59/136)}
& 30.1 {\footnotesize (41/136)} \\

$P(\mathrm{Branch}\geq80\% \mid \mathrm{MRC}_{\theta})$
& 71.6 {\footnotesize (83/116)}
& 67.8 {\footnotesize (59/87)}
& 64.1 {\footnotesize (41/64)} \\
\bottomrule
\end{tabular}
\end{table}

\begin{tcolorbox}[colback=gray!5, colframe=gray!30, title=\textbf{\small Answer to RQ3}, coltitle=black!90, left=2.5pt, top=3pt, bottom=3pt, right=2.5pt, boxsep=1.8pt]
MR coverage provides adequacy information beyond execution-based coverage. Although MR metrics increase with sampled test-suite size, weak within-size correlations show that high statement or branch coverage does not necessarily imply high behavioral validation. \textit{MR Cover} specifically captures whether exercised behaviors are explicitly checked against inferred behavioral relations.
\end{tcolorbox}

\subsection{RQ4: Practical Relevance}\label{sec:rq4-results}

RQ4 examines the practical relevance of the inferred MR space and MR-based coverage labels. We evaluate this through three complementary analyses: issue-description mapping, targeted assertion augmentation for weak-oracle relations, and MR-relevant injected fault detection. These analyses are intended to provide supporting evidence rather than causal evidence of real-world fault proneness.

\subsubsection{Reported issue annotations are substantially represented in the inferred MR space.}


\begin{table}[h]
\centering
\caption{Summary of reported issue annotations mapped to the inferred MR space.}
\label{tab:issue-mr-summary}
\small
\setlength{\tabcolsep}{4pt}
\begin{tabular}{lrr}
\toprule
\textbf{Outcome} & \textbf{Count} & \textbf{\%} \\
\midrule
Issue annotations & 634 & 100.0 \\
Mapped to MR relation types & 459 & 72.4 \\
\quad Direct mapping & 90 & 14.2 \\
\quad Inferred mapping & 262 & 41.3 \\
\quad Multiple MR mapping & 107 & 16.9 \\
No clear MR match & 175 & 27.6 \\
\bottomrule
\end{tabular}
\end{table}

We first examine whether reported issue descriptions can be related to the inferred MR space. We collected component-related issue reports and manually annotated 634 behavioral descriptions, referred to as issue annotations. Table~\ref{tab:issue-mr-summary} shows that, under a strict criterion requiring a unique MR match, 352 of 634 issue annotations (55.5\%) align with inferred MR relation types. Including annotations with multiple MR matches increases the mapped set to 459 annotations (72.4\%), while the remaining 175 annotations (27.6\%) have no clear match in the current MR space.

\begin{figure}[h]
  \centering
  \includegraphics[width=0.84\linewidth]{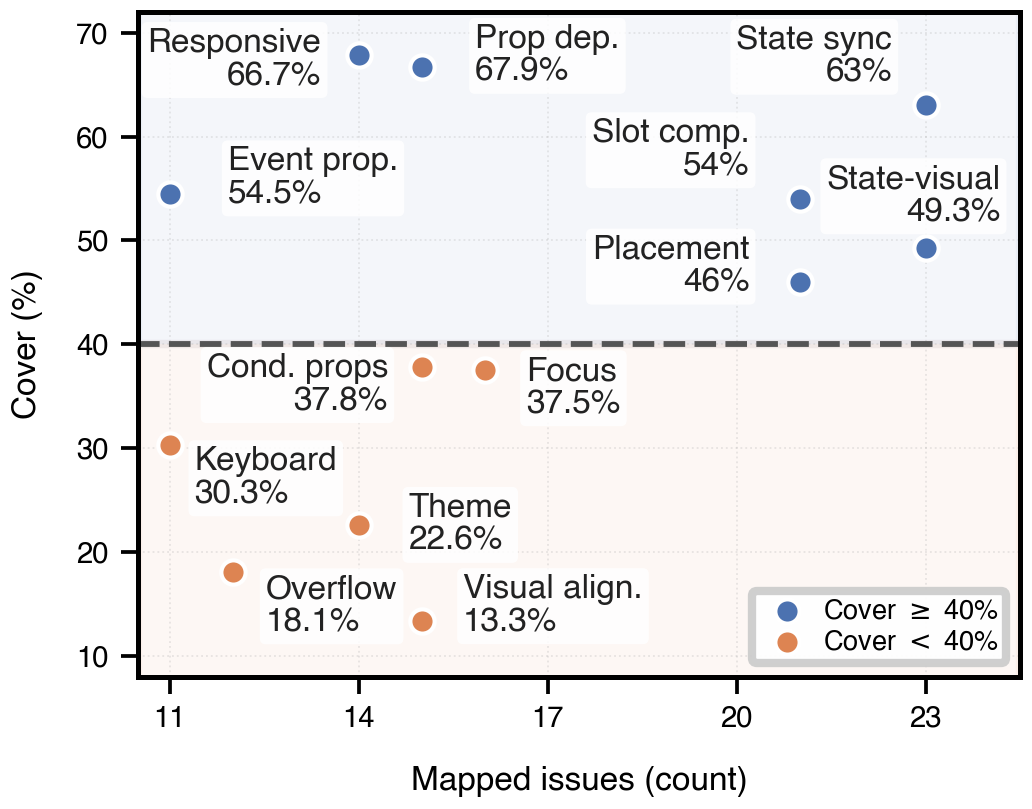}
  \caption{Relationship between mapped issue annotations and average MR Cover across MR relation types. The dashed line at 40\% is a visual reference for lower-cover relation types, not a decision threshold.}
  \label{fig:rq4_issue_scatter}
\end{figure}

Figure~\ref{fig:rq4_issue_scatter} further relates mapped issue annotations to average MR \textit{Cover} across relation types. The mapped issues are distributed across both higher and lower-cover relation types. Frequently mapped relations such as state synchronization and state--visual mapping show relatively high \textit{Cover}, whereas visual alignment, overflow handling, theme consistency, and keyboard interaction fall below the 40\% visual reference line. These results suggest that the inferred MR space represents many behavioral concerns described in reported issues, and that MR \textit{Cover} helps indicate which issue-aligned areas remain less explicitly validated.

\subsubsection{Weak-oracle relations highlight opportunities for additional behavioral validation.}

As a second perspective on practical relevance, we conducted the oracle-strengthening study described in Section~\ref{sec:external-validation} to examine whether weak-oracle relations correspond to opportunities for strengthening existing tests through targeted assertions derived from the inferred MRs.

\begin{table}[h]
\centering
\caption{Weak-oracle assertion study summary. Percentages are computed with respect to the denominator shown in each block.}
\label{tab:wo-study}
\setlength{\tabcolsep}{3pt}
\begin{tabular}{@{}lcc@{}}
\toprule
\textbf{Outcome} & \textbf{Count} & \textbf{\%} \\
\midrule
\multicolumn{3}{@{}l}{\textbf{Sampled weak-oracle relations} ($n=200$)} \\
\quad Executed with augmented assertion & 200 & 100.0 \\
\midrule
\multicolumn{3}{@{}l}{\textbf{Assertion result} ($n=200$)} \\
\quad Passed -- assertion satisfied & 81 & 40.5 \\
\quad Failed -- assertion not satisfied & 119 & 59.5 \\
\midrule
\multicolumn{3}{@{}l}{\textbf{Failure inspection} ($n=119$)} \\
\quad Mismatch with inferred MR assertion & 84 & 70.6 \\
\quad Snapshot artifact & 14 & 11.8 \\
\quad Indeterminate & 21 & 17.6 \\
\midrule
\textbf{Conclusive outcomes} ($81+84$ of 200) & \textbf{165} & \textbf{82.5} \\
\bottomrule
\end{tabular}
\end{table}

Table~\ref{tab:wo-study} summarizes the results. All 200 augmented tests executed successfully. Among them, 81 added assertions passed, whereas 119 failed. Manual inspection attributed 84 of the failed cases to mismatches between the observed behavior and the inferred MR assertion; the remaining failures resulted from snapshot artifacts or indeterminate execution conditions. These mismatches should not be interpreted as confirmed defects or as direct evidence that the inferred MR is correct. Rather, they identify cases where the existing tests do not provide sufficient oracle evidence to determine whether the inferred relation holds.
Overall, 165 cases (82.5\%) produced informative outcomes, either because the added assertion was satisfied or because manual inspection identified an MR-related mismatch. These results suggest that many weak-oracle relations correspond to behaviors that are exercised but not explicitly validated by existing tests. More broadly, targeted assertion augmentation provides additional behavioral evidence for identifying opportunities to strengthen behavioral validation in existing tests.

\subsubsection{MR labels show a trend in behavior-oriented fault detection.}
To further examine this trend in our targeted injected-fault sample, we conducted the injected-fault study described in Section~\ref{sec:external-validation}. This study is not intended to estimate general fault-detection capability; rather, it provides a focused check of whether MR labels are consistent with detection outcomes for faults explicitly constructed to violate the corresponding MR.
Table~\ref{tab:mr_fault_detection} shows a consistent gradient trend across MR labels. Tests associated with \textit{Covered} relations detected 48 of 80 injected faults (60.0\%), compared with 38 of 80 (47.5\%) for \textit{Weak} relations, whereas none of the 40 faults associated with \textit{Untouched} relations were detected. The zero detection rate for \textit{Untouched} relations is expected: by definition, these relations correspond to behaviors that are not exercised by the original test suites. We include this group as a sanity-check baseline, rather than as evidence of relative oracle strength.

\begin{table}[h]
\centering
\caption{Detection of MR-relevant injected faults grouped by MR status.}
\label{tab:mr_fault_detection}
\small
\setlength{\tabcolsep}{4pt}
\begin{tabular}{lrrr}
\toprule
\textbf{MR Status} & \textbf{\#Faults} & \textbf{\#Detected} & \textbf{Kill Rate} (\%) \\
\midrule
Covered   & 80 & 48 & 60.0 \\
Weak      & 80 & 38 & 47.5 \\
Untouched & 40 &  0 & 0.0 \\
\bottomrule
\end{tabular}
\end{table}

This pattern is consistent with the intended interpretation of MR labels: relations with explicit validation provide stronger protection against MR-relevant injected faults than relations that are only exercised, while unexercised relations provide no detection evidence in this study. A representative pair illustrates this distinction. A \textit{Covered} MR in the \textit{Infinite Scroll} component detected a fault replacing \textit{delay} with \textit{9999}, while a \textit{Weak} MR in the \textit{Space} component left all tests passing after changing \textit{wrap=false} to \textit{wrap=undefined}. Overall, these results provide supporting evidence that MR labels are consistent with differences in detecting MR-relevant behavioral faults in this targeted sample.

\begin{tcolorbox}[colback=gray!5, colframe=gray!30, title=\textbf{\small Answer to RQ4}, coltitle=black!90, left=2.5pt, top=3pt, bottom=3pt, right=2.5pt, boxsep=1.8pt]
Reported issue descriptions are substantially represented in the inferred MR space. Weak-oracle relations often identify exercised behaviors with limited oracle evidence, and MR labels show a consistent trend in the targeted injected-fault study. Overall, these results support the practical relevance of MR coverage as a behavioral validation signal.
\end{tcolorbox}

\section{Discussion}\label{sec:discussion}
\noindent\textbf{MR Coverage Complements Execution-based Coverage.}
Our results suggest that MR coverage captures a dimension of test adequacy that is not reflected by execution-based coverage alone. While statement and branch coverage indicate whether implementation code is exercised, MR coverage considers whether behavioral relations in the inferred MR space are exercised and explicitly validated. These two perspectives therefore provide complementary information for assessing existing UI component test suites.

\noindent\textbf{Interpreting Behavioral Gaps.}
The distinction between \textit{Touch} and \textit{Cover} provides additional context for interpreting behavioral gaps within the inferred MR space. Low \textit{Touch} may indicate relations that are not exercised by existing tests, whereas high \textit{Touch} but low \textit{Cover} suggests that the corresponding behaviors are exercised without explicit relation-relevant validation. Such distinctions may help prioritize where additional behavioral assertions are likely to be beneficial.

\noindent\textbf{Using Inferred MRs as a Behavioral Reference.}
The inferred MR space is not intended to serve as a complete behavioral specification. Instead, it provides an empirical behavioral reference for aggregate adequacy assessment. The manual validation, issue-mapping analysis, oracle-strengthening study, and MR-relevant injected-fault study together provide supporting evidence that the inferred MR space captures many practically relevant behavioral relations, while also leaving behaviors outside the current taxonomy and inference scope.
\section{Threats to Validity}\label{sec:threats_to_validity}

\noindent\textbf{Construct validity.}
MR coverage provides a relation-level adequacy signal over an inferred MR space rather than a complete behavioral specification. Because this space is inferred from public artifacts, it may omit intended behaviors, include imperfect relations, or depend on the granularity at which behavioral relations are represented. For example, a broad MR may be decomposed into multiple finer-grained relations, leading to different \textit{Touch} and \textit{Cover} values. We mitigate these risks through taxonomy-constrained inference, multi-artifact grounding, hybrid deterministic--semantic alignment, and manual validation of sampled MRs and alignment decisions. The oracle-strengthening study introduces an additional threat because injected assertions only approximate inferred relations; we reduce this risk through relation-specific templates and manual inspection of failing executions. MR coverage should therefore be interpreted with respect to the inferred MR space rather than as an absolute measure of behavioral completeness.

\vspace{4pt}
\noindent\textbf{Internal validity.}
MR inference and semantic alignment rely partly on LLM reasoning and may be affected by model-specific biases. We mitigate this threat through fixed prompts, fixed inference parameters, a fixed taxonomy, identical analysis procedures, and evaluation across multiple LLMs. Although the inferred MR sets differ across models, the main findings remain consistent across the evaluated configurations.

\vspace{4pt}
\noindent\textbf{External validity.}
Our evaluation covers four mature open-source UI component libraries spanning the React and Vue ecosystems. The findings may not generalize to proprietary frameworks or application-specific UI components. In addition, the oracle-strengthening study was conducted on two libraries with reproducible execution environments, and different libraries or assertion templates may lead to different quantitative results. The findings should therefore be interpreted within the studied setting, although the proposed assessment framework is not tied to a specific UI library or framework.
\section{Related work}\label{sec:related_work}

\noindent\textbf{Execution-based Coverage and Test Adequacy.}
Statement and branch coverage remain the dominant measures of test adequacy, yet their relationship with fault-detection effectiveness is known to be limited~\cite{papadakis2019mutation, soremekun2023evaluating,inozemtseva2014coverage,gopinath2014code}. Test suites with similar execution coverage may exhibit substantially different defect-detection capabilities, especially for UI component libraries where correctness depends on behavioral relations among component properties, interactions, and rendering outcomes. Our work complements execution-based coverage by assessing whether inferred behavioral relations are exercised and explicitly validated.

\vspace{2pt}
\noindent\textbf{Metamorphic Testing and MR Inference.}
Metamorphic Testing (MT) addresses the oracle problem by specifying expected relations between multiple executions rather than absolute outputs~\cite{chen2020metamorphic, segura2016survey, liu2013effectively}. A long-standing challenge is the construction of meaningful metamorphic relations~\cite{chen2018metamorphic}. Recent studies have shown that large language models can effectively infer MRs from source code, documentation, and natural-language artifacts~\cite{NolascoMDGGPUAF24,zhang2023automated,shin2024towards, ayerdi2024genmorph, duque2024selecting, le2025can, ma2025specgen}. Existing work primarily uses inferred MRs for follow-up test generation or oracle construction. In contrast, we use inferred MRs as a behavioral reference for assessing the adequacy of existing UI component test suites.

\vspace{2pt}
\noindent\textbf{GUI and Component Testing.}
Prior work on GUI testing has focused on model-based, event-driven, and automated exploration techniques, with particular attention to the oracle problem~\cite{memon2001hierarchical,xie2007designing,barr2014oracle,wang2020combodroid,wang2021vet,ahlgren2021testing}. Existing studies of UI component testing mainly evaluate test quality through structural metrics, test smells, or co-evolution analyses rather than behavioral specifications~\cite{spadini2018relation,rwemalika2023smells}. Our work instead introduces a relation-level adequacy perspective based on inferred MRs.

\vspace{2pt}
\noindent\textbf{Relation to Other Adequacy Techniques.}
While MR coverage builds on metamorphic relations, its objective differs from mutation testing, checked coverage, and property-based testing. Rather than generating tests, injecting faults, or measuring assertion reachability, it uses inferred behavioral relations as a reference for assessing the adequacy of existing UI component test suites. Thus, MR coverage provides a complementary relation-level perspective on behavioral validation.
\section{Conclusion}\label{sec:conclusion}

This paper presented an MR-based framework for assessing behavioral validation in UI component test suites. The framework distinguishes behavioral reach (\textit{Touch}) from behavioral validation (\textit{Cover}) through a UI-specific MR taxonomy, LLM-assisted MR inference, and hybrid test--MR alignment.
Our evaluation on 214 components from four production UI libraries shows that existing test suites exercise substantially more behavioral relations than they explicitly validate. These differences are not reflected by statement or branch coverage alone, indicating that execution-based coverage provides only a partial view of behavioral adequacy. Additional analyses based on reported issue descriptions, oracle strengthening, and MR-relevant injected faults further support the practical relevance of the inferred MR space and MR coverage.
Overall, MR coverage provides a relation-level adequacy signal using inferred metamorphic relations as an empirical behavioral reference. It complements execution-based coverage by providing additional insight into behavioral validation in modern UI component test suites.



\bibliographystyle{IEEEtran}
\bibliography{references}

@article{chen2018metamorphic,
  title={Metamorphic testing: A review of challenges and opportunities},
  author={Chen, Tsong Yueh and Kuo, Fei-Ching and Liu, Huai and Poon, Pak-Lok and Towey, Dave and Tse, TH and Zhou, Zhi Quan},
  journal={ACM Computing Surveys (CSUR)},
  volume={51},
  number={1},
  pages={1--27},
  year={2018},
  publisher={ACM New York, NY, USA}
}

@article{segura2016survey,
  title={A survey on metamorphic testing},
  author={Segura, Sergio and Fraser, Gordon and Sanchez, Ana B and Ruiz-Cort{\'e}s, Antonio},
  journal={IEEE Transactions on software engineering},
  volume={42},
  number={9},
  pages={805--824},
  year={2016},
  publisher={IEEE}
}

@inproceedings{inozemtseva2014coverage,
  title={Coverage is not strongly correlated with test suite effectiveness},
  author={Inozemtseva, Laura and Holmes, Reid},
  booktitle={Proceedings of the 36th international conference on software engineering},
  pages={435--445},
  year={2014}
}

@article{kochhar2017code,
  title={Code coverage and postrelease defects: A large-scale study on open source projects},
  author={Kochhar, Pavneet Singh and Lo, David and Lawall, Julia and Nagappan, Nachiappan},
  journal={IEEE Transactions on Reliability},
  volume={66},
  number={4},
  pages={1213--1228},
  year={2017},
  publisher={IEEE}
}

@inproceedings{andrews2005mutation,
  title={Is mutation an appropriate tool for testing experiments?},
  author={Andrews, James H and Briand, Lionel C and Labiche, Yvan},
  booktitle={Proceedings of the 27th international conference on Software engineering},
  pages={402--411},
  year={2005}
}

@inproceedings{tian2018deeptest,
  title={Deeptest: Automated testing of deep-neural-network-driven autonomous cars},
  author={Tian, Yuchi and Pei, Kexin and Jana, Suman and Ray, Baishakhi},
  booktitle={Proceedings of the 40th international conference on software engineering},
  pages={303--314},
  year={2018}
}

@inproceedings{zhang2018deeproad,
  title={Deeproad: Gan-based metamorphic testing and input validation framework for autonomous driving systems},
  author={Zhang, Mengshi and Zhang, Yuqun and Zhang, Lingming and Liu, Cong and Khurshid, Sarfraz},
  booktitle={Proceedings of the 33rd ACM/IEEE international conference on automated software engineering},
  pages={132--142},
  year={2018}
}

@inproceedings{chen2016empirical,
  title={An empirical comparison of compiler testing techniques},
  author={Chen, Junjie and Hu, Wenxiang and Hao, Dan and Xiong, Yingfei and Zhang, Hongyu and Zhang, Lu and Xie, Bing},
  booktitle={Proceedings of the 38th International Conference on Software Engineering},
  pages={180--190},
  year={2016}
}

@inproceedings{gopinath2014code,
  title={Code coverage for suite evaluation by developers},
  author={Gopinath, Rahul and Jensen, Carlos and Groce, Alex},
  booktitle={Proceedings of the 36th international conference on software engineering},
  pages={72--82},
  year={2014}
}

@article{chen2020metamorphic,
  title={Metamorphic testing: a new approach for generating next test cases},
  author={Chen, Tsong Y and Cheung, Shing C and Yiu, Shiu Ming},
  journal={arXiv preprint arXiv:2002.12543},
  year={2020}
}

@inproceedings{zhang2023automated,
  title={Automated metamorphic-relation generation with ChatGPT: An experience report},
  author={Zhang, Yifan and Towey, Dave and Pike, Matthew},
  booktitle={2023 IEEE 47th Annual Computers, Software, and Applications Conference (COMPSAC)},
  pages={1780--1785},
  year={2023},
  organization={IEEE}
}

@inproceedings{shin2024towards,
  title={Towards generating executable metamorphic relations using large language models},
  author={Shin, Seung Yeob and Pastore, Fabrizio and Bianculli, Domenico and Baicoianu, Alexandra},
  booktitle={International Conference on the Quality of Information and Communications Technology},
  pages={126--141},
  year={2024},
  organization={Springer}
}

@article{memon2001hierarchical,
  title={Hierarchical GUI test case generation using automated planning},
  author={Memon, Atif M and Pollack, Martha E and Soffa, Mary Lou},
  journal={IEEE transactions on software engineering},
  volume={27},
  number={2},
  pages={144--155},
  year={2001},
  publisher={IEEE}
}

@article{xie2007designing,
  title={Designing and comparing automated test oracles for GUI-based software applications},
  author={Xie, Qing and Memon, Atif M},
  journal={ACM Transactions on Software Engineering and Methodology (TOSEM)},
  volume={16},
  number={1},
  pages={4--es},
  year={2007},
  publisher={ACM New York, NY, USA}
}

@article{barr2014oracle,
  title={The oracle problem in software testing: A survey},
  author={Barr, Earl T and Harman, Mark and McMinn, Phil and Shahbaz, Muzammil and Yoo, Shin},
  journal={IEEE transactions on software engineering},
  volume={41},
  number={5},
  pages={507--525},
  year={2014},
  publisher={IEEE}
}

@inproceedings{wang2020combodroid,
  title={Combodroid: generating high-quality test inputs for android apps via use case combinations},
  author={Wang, Jue and Jiang, Yanyan and Xu, Chang and Cao, Chun and Ma, Xiaoxing and Lu, Jian},
  booktitle={Proceedings of the ACM/IEEE 42nd International Conference on Software Engineering},
  pages={469--480},
  year={2020}
}

@inproceedings{wang2021vet,
  title={Vet: identifying and avoiding UI exploration tarpits},
  author={Wang, Wenyu and Yang, Wei and Xu, Tianyin and Xie, Tao},
  booktitle={Proceedings of the 29th ACM Joint Meeting on European Software Engineering Conference and Symposium on the Foundations of Software Engineering},
  pages={83--94},
  year={2021}
}

@inproceedings{ahlgren2021testing,
  title={Testing web enabled simulation at scale using metamorphic testing},
  author={Ahlgren, John and Berezin, Maria and Bojarczuk, Kinga and Dulskyte, Elena and Dvortsova, Inna and George, Johann and Gucevska, Natalija and Harman, Mark and Lomeli, Maria and Meijer, Erik and others},
  booktitle={2021 IEEE/ACM 43rd International Conference on Software Engineering: Software Engineering in Practice (ICSE-SEIP)},
  pages={140--149},
  year={2021},
  organization={IEEE}
}

@inproceedings{spadini2018relation,
  title={On the relation of test smells to software code quality},
  author={Spadini, Davide and Palomba, Fabio and Zaidman, Andy and Bruntink, Magiel and Bacchelli, Alberto},
  booktitle={2018 IEEE international conference on software maintenance and evolution (ICSME)},
  pages={1--12},
  year={2018},
  organization={IEEE Computer Society}
}

@article{rwemalika2023smells,
  title={Smells in system user interactive tests},
  author={Rwemalika, Renaud and Habchi, Sarra and Papadakis, Mike and Le Traon, Yves and Brasseur, Marie-Claude},
  journal={Empirical Software Engineering},
  volume={28},
  number={1},
  pages={20},
  year={2023},
  publisher={Springer}
}

@article{delgado2016reusing,
  title={Reusing UI elements with model-based user interface development},
  author={Delgado, Antonio and Estepa, Antonio and Troyano, JA and Estepa, Rafael},
  journal={International Journal of Human-Computer Studies},
  volume={86},
  pages={48--62},
  year={2016},
  publisher={Elsevier}
}

@inproceedings{abdalkareem2017developers,
  title={Why do developers use trivial packages? an empirical case study on npm},
  author={Abdalkareem, Rabe and Nourry, Olivier and Wehaibi, Sultan and Mujahid, Suhaib and Shihab, Emad},
  booktitle={Proceedings of the 2017 11th joint meeting on foundations of software engineering},
  pages={385--395},
  year={2017}
}

@article{decan2019empirical,
  title={An empirical comparison of dependency network evolution in seven software packaging ecosystems},
  author={Decan, Alexandre and Mens, Tom and Grosjean, Philippe},
  journal={Empirical Software Engineering},
  volume={24},
  number={1},
  pages={381--416},
  year={2019},
  publisher={Springer}
}

@article{weeraddana2024dependency,
  title={Dependency-induced waste in continuous integration: An empirical study of unused dependencies in the npm ecosystem},
  author={Weeraddana, Nimmi Rashinika and Alfadel, Mahmoud and McIntosh, Shane},
  journal={Proceedings of the ACM on Software Engineering},
  volume={1},
  number={FSE},
  pages={2632--2655},
  year={2024},
  publisher={ACM New York, NY, USA}
}

@article{venturini2023depended,
  title={I depended on you and you broke me: An empirical study of manifesting breaking changes in client packages},
  author={Venturini, Daniel and Cogo, Filipe Roseiro and Polato, Ivanilton and Gerosa, Marco A and Wiese, Igor Scaliante},
  journal={ACM Transactions on Software Engineering and Methodology},
  volume={32},
  number={4},
  pages={1--26},
  year={2023},
  publisher={ACM New York, NY, USA}
}

@inproceedings{alshayban2020accessibility,
  title={Accessibility issues in android apps: state of affairs, sentiments, and ways forward},
  author={Alshayban, Abdulaziz and Ahmed, Iftekhar and Malek, Sam},
  booktitle={Proceedings of the ACM/IEEE 42nd International Conference on Software Engineering},
  pages={1323--1334},
  year={2020}
}

@article{lazuardy2022modern,
  title={Modern front end web architectures with react. js and next. js},
  author={Lazuardy, Mochammad Fariz Syah and Anggraini, Dyah},
  journal={Research Journal of Advanced Engineering and Science},
  volume={7},
  number={1},
  pages={132--141},
  year={2022}
}

@article{naik2023awesome,
  title={Awesome React. js (Unleash the Power of Modern UI Building)},
  author={Naik, Poornima G and Oza, Kavita},
  journal={International Institute of Organized Research (I2OR)},
  year={2023}
}

@article{zhou2025declarui,
  title={Declarui: Bridging design and development with automated declarative ui code generation},
  author={Zhou, Ting and Zhao, Yanjie and Hou, Xinyi and Sun, Xiaoyu and Chen, Kai and Wang, Haoyu},
  journal={Proceedings of the ACM on Software Engineering},
  volume={2},
  number={FSE},
  pages={219--241},
  year={2025},
  publisher={ACM New York, NY, USA}
}

@inproceedings{cho2025metamorphic,
  title={Metamorphic testing of large language models for natural language processing},
  author={Cho, Steven and Ruberto, Stefano and Terragni, Valerio},
  booktitle={2025 IEEE International Conference on Software Maintenance and Evolution (ICSME)},
  pages={174--186},
  year={2025},
  organization={IEEE}
}

@inproceedings{tsigkanos2023large,
  title={Large language models: The next frontier for variable discovery within metamorphic testing?},
  author={Tsigkanos, Christos and Rani, Pooja and M{\"u}ller, Sebastian and Kehrer, Timo},
  booktitle={2023 IEEE International Conference on Software Analysis, Evolution and Reengineering (SANER)},
  pages={678--682},
  year={2023},
  organization={IEEE}
}

@article{jartarghar2022react,
  title={React apps with Server-Side rendering: Next. js},
  author={Jartarghar, Harish A and Salanke, Girish Rao and AR, Ashok Kumar and GS, Sharvani and Dalali, Shivakumar},
  journal={Journal of Telecommunication, Electronic and Computer Engineering (JTEC)},
  volume={14},
  number={4},
  pages={25--29},
  year={2022}
}

@inproceedings{lu2025misty,
  title={Misty: Ui prototyping through interactive conceptual blending},
  author={Lu, Yuwen and Leung, Alan and Swearngin, Amanda and Nichols, Jeffrey and Barik, Titus},
  booktitle={Proceedings of the 2025 CHI Conference on Human Factors in Computing Systems},
  pages={1--17},
  year={2025}
}

@article{liu2024enhancing,
  title={Enhancing user engagement through adaptive UI/UX design: A study on personalized mobile app interfaces},
  author={Liu, Yingchia and Tan, Hao and Cao, Guanghe and Xu, Yang},
  journal={Computer Science \& IT Research Journal},
  volume={5},
  number={8},
  pages={1942--1962},
  year={2024}
}

@inproceedings{zhang2024llamatouch,
  title={Llamatouch: A faithful and scalable testbed for mobile ui task automation},
  author={Zhang, Li and Wang, Shihe and Jia, Xianqing and Zheng, Zhihan and Yan, Yunhe and Gao, Longxi and Li, Yuanchun and Xu, Mengwei},
  booktitle={Proceedings of the 37th Annual ACM Symposium on User Interface Software and Technology},
  pages={1--13},
  year={2024}
}

@inproceedings{soremekun2023evaluating,
  title={Evaluating the impact of experimental assumptions in automated fault localization},
  author={Soremekun, Ezekiel and Kirschner, Lukas and B{\"o}hme, Marcel and Papadakis, Mike},
  booktitle={2023 IEEE/ACM 45th International Conference on Software Engineering (ICSE)},
  pages={159--171},
  year={2023},
  organization={IEEE}
}

@inproceedings{huo2014improving,
  title={Improving oracle quality by detecting brittle assertions and unused inputs in tests},
  author={Huo, Chen and Clause, James},
  booktitle={Proceedings of the 22nd ACM SIGSOFT International Symposium on Foundations of Software Engineering},
  pages={621--631},
  year={2014}
}

@inproceedings{schuler2011assessing,
  title={Assessing oracle quality with checked coverage},
  author={Schuler, David and Zeller, Andreas},
  booktitle={2011 Fourth IEEE International Conference on Software Testing, Verification and Validation},
  pages={90--99},
  year={2011},
  organization={IEEE}
}

@article{zhou2018metamorphic,
  title={Metamorphic relations for enhancing system understanding and use},
  author={Zhou, Zhi Quan and Sun, Liqun and Chen, Tsong Yueh and Towey, Dave},
  journal={IEEE Transactions on Software Engineering},
  volume={46},
  number={10},
  pages={1120--1154},
  year={2018},
  publisher={IEEE}
}

@article{li2025metamorphic,
  title={Metamorphic relation generation: State of the art and research directions},
  author={Li, Rui and Liu, Huai and Poon, Pak-Lok and Towey, Dave and Sun, Chang-Ai and Zheng, Zheng and Zhou, Zhi Quan and Chen, Tsong Yueh},
  journal={ACM Transactions on Software Engineering and Methodology},
  volume={34},
  number={5},
  pages={1--25},
  year={2025},
  publisher={ACM New York, NY}
}

@article{ayerdi2024genmorph,
  title={Genmorph: Automatically generating metamorphic relations via genetic programming},
  author={Ayerdi, Jon and Terragni, Valerio and Jahangirova, Gunel and Arrieta, Aitor and Tonella, Paolo},
  journal={IEEE Transactions on Software Engineering},
  volume={50},
  number={7},
  pages={1888--1900},
  year={2024},
  publisher={IEEE}
}

@inproceedings{duque2024selecting,
  title={Selecting and constraining metamorphic relations},
  author={Duque-Torres, Alejandra},
  booktitle={Proceedings of the 2024 IEEE/ACM 46th International Conference on Software Engineering: Companion Proceedings},
  pages={212--216},
  year={2024}
}

@article{liu2013effectively,
  title={How effectively does metamorphic testing alleviate the oracle problem?},
  author={Liu, Huai and Kuo, Fei-Ching and Towey, Dave and Chen, Tsong Yueh},
  journal={IEEE Transactions on Software Engineering},
  volume={40},
  number={1},
  pages={4--22},
  year={2013},
  publisher={IEEE}
}

@incollection{papadakis2019mutation,
  title={Mutation testing advances: an analysis and survey},
  author={Papadakis, Mike and Kintis, Marinos and Zhang, Jie and Jia, Yue and Le Traon, Yves and Harman, Mark},
  booktitle={Advances in computers},
  volume={112},
  pages={275--378},
  year={2019},
  publisher={Elsevier}
}

@inproceedings{goldstein2024property,
  title={Property-based testing in practice},
  author={Goldstein, Harrison and Cutler, Joseph W and Dickstein, Daniel and Pierce, Benjamin C and Head, Andrew},
  booktitle={Proceedings of the IEEE/ACM 46th International Conference on Software Engineering},
  pages={1--13},
  year={2024}
}

@inproceedings{bajammal2021semantic,
  title={Semantic web accessibility testing via hierarchical visual analysis},
  author={Bajammal, Mohammad and Mesbah, Ali},
  booktitle={2021 IEEE/ACM 43rd International Conference on Software Engineering (ICSE)},
  pages={1610--1621},
  year={2021},
  organization={IEEE}
}

@inproceedings{le2025can,
  title={Can LLMs reason about program semantics? a comprehensive evaluation of LLMs on formal specification inference},
  author={Le-Cong, Thanh and Le, Bach and Murray, Toby},
  booktitle={Proceedings of the 63rd Annual Meeting of the Association for Computational Linguistics (Volume 1: Long Papers)},
  pages={21991--22014},
  year={2025}
}

@inproceedings{ma2025specgen,
  title={Specgen: Automated generation of formal program specifications via large language models},
  author={Ma, Lezhi and Liu, Shangqing and Li, Yi and Xie, Xiaofei and Bu, Lei},
  booktitle={2025 IEEE/ACM 47th International Conference on Software Engineering (ICSE)},
  pages={16--28},
  year={2025},
  organization={IEEE}
}

@article{NolascoMDGGPUAF24,
  author       = {Agust{\'{\i}}n Nolasco and
                  Facundo Molina and
                  Renzo Degiovanni and
                  Alessandra Gorla and
                  Diego Garbervetsky and
                  Mike Papadakis and
                  Sebasti{\'{a}}n Uchitel and
                  Nazareno Aguirre and
                  Marcelo F. Frias},
  title        = {Abstraction-Aware Inference of Metamorphic Relations},
  journal      = {Proc. {ACM} Softw. Eng.},
  volume       = {1},
  number       = {{FSE}},
  pages        = {450--472},
  year         = {2024},
  url          = {https://doi.org/10.1145/3643747},
  doi          = {10.1145/3643747},
}

\end{document}